\documentstyle[draft,prl,aps,twocolumn]{revtex}
\tighten
\begin{document}
\title{Inflation and the Fine-Tuning Problem}
\author{ Andrew Matacz
\thanks{e-mail address: andrewm@maths.su.oz.au}}
\address{School of Mathematics and Statistics \\
University of Sydney\\ NSW, 2006, Australia}
\maketitle
\begin{abstract}
I describe a recently derived stochastic approach to inflaton dynamics 
which can address some serious problems associated with conventional 
inflationary theory. 
Using this theory I derive an exact solution to the stochastic dynamics 
for the case of a $\lambda\phi^4$ potential and use it to study the 
generated primordial 
density fluctuations. It is found that on both sub and super-horizon scales the
theory predicts gaussian fluctuations to a very high accuracy along with a 
near scale-invariant spectrum. Of most interest is that the amplitude 
constraint is found to be satisfied for $\lambda\sim 10^{-5}$ rather than 
$\lambda\sim 10^{-14}$ of the conventional theory. This represents a 
dramatic easing of the fine-tuning constraints, a feature likely 
to generalize to a wide range of potentials.  
\end{abstract}

\vspace{1cm}
The inflationary universe scenario asserts that, at some very early
time,
the universe went through a de Sitter phase expansion with scale
factor $a(t)$ growing as $e^{Ht}$. Inflation is
needed because it solves the horizon, flatness and monopole problems
of the very early universe
and also provides a mechanism for the creation of
primordial density fluctuations. For these reasons it is an integral
part of the standard cosmological model
\cite{inflation}.

The inflationary phase is driven by a quantum
scalar field with a potential $V(\Phi)$, that can take on many
different
forms that satisfy the `slow roll' conditions. 
In the conventional approach to inflaton dynamics \cite{inflation},
the inflaton field $\Phi$ is first
split into a spatially homogeneous piece and an inhomogeneous piece
\begin{equation}
\Phi({\bf x},s)=\phi(s)+\psi({\bf x},s).
\end{equation}
The dynamics of the $\phi$ is then postulated to
obey the classical `slow roll' equation of motion
\begin{equation}
\dot{\phi}+\frac{V'(\phi)}{3H}=0.
\end{equation}
This equation governs the dynamics of $\phi$ which drives the
inflationary phase.
It is also possible to discuss the generation of primordial density
fluctuations using $\psi$.
Assuming that $\phi\gg \psi$,
it can be shown that $\psi$ is described by a free massless
minimally coupled quantum scalar field. During exponential inflation the
quantum fluctuations
of $\psi$ grow as \cite{vfl}
\begin{equation}
\langle\psi^2\rangle=(2\pi)^{-2} H^3t.
\end{equation}
These quantum
fluctuations are then identified with the classical
fluctuations which generate primordial density fluctuations
\cite{inflation,fluct}. It is important to note here that interactions 
between the coarse-grained field $\phi$ and its fluctuations 
$\psi$ are ignored. This is possible in this approach 
because the density fluctuations are directly identified 
with $\langle\psi^2\rangle$. 

Consistent with the conventional approach above is the `Stochastic Inflation' 
program initiated by Starobinsky \cite{staro} and further developed 
by others \cite{stoinf}. In this case the field $\phi$ obeys
\begin{equation}
\dot \phi + \frac{V'(\phi)}{3H} = \frac{H^{3/2}}{2\pi}F_w (t),
\end{equation}
where $F_w(t)$ is a zero mean gaussian white noise source of unit
amplitude. 
In this case $\phi$ describes the field $\Phi$ coarse-grained 
over a volume determined by the de Sitter Hubble radius. We 
will refer to $\phi$ as the local order parameter. In this method the 
observable universe is comprised of many patches each with its own local order 
parameter whose dynamics obeys (4). Spatial inhomogeneities arise because 
the local order parameter in each patch can take on different values by virtue 
of the noise in (4). Equation (4) has been the basis for many applications 
including studies of the generation of primordial density
fluctuations \cite{sifluct,yvm} and the very large scale
structure of the universe \cite{vlss}.

A problem with the conventional approach is that it is assumed, without 
justification, that the local order parameter $\phi$ can be treated 
as a classical order parameter, and that the quantum fluctuations of
$\psi$ are equivalent to classical fluctuations. Since $\phi$ and 
$\psi$ are treated as de-coupled closed quantum systems it is impossible
for this method to explain the quantum-to-classical transition of $\phi$ 
and $\psi$ from first principles. Another more serious problem comes from
directly identifying the quantum fluctuations $\langle\psi^2\rangle$, 
with the classical fluctuations that generate primordial density fluctuations. 
This scheme leads to an overproduction of
primordial density fluctuations which can
only be avoided by unnaturally fine-tuning the coupling constants in
the inflaton potential. This is the well known fine-tuning problem 
of inflation. 

Several authors have previously suggested that 
these problems arise because the conventional approach to calculating
primordial density fluctuations is inconsistent with the established 
methods of non-equilibrium statistical physics. This was first pointed 
out by Hu and Zhang \cite{cgea} and further developed in 
\cite{HuBelgium} (see also Lombardo and Mazzitelli 
\cite{sim} and Morikawa \cite{mm}). Morikawa \cite{morwas} 
first suggested that this inconsistency was the origin of the fine-tuning 
problem. Calzetta and Hu \cite{CH1} and more recently Calzetta and 
Gonorazky \cite{calgon} independently and in much greater detail 
addressed this issue for a $\lambda\phi^4$ theory.

While the conventional approach may be the only possible one 
for a free field, in \cite{matacz} an alternative has been developed  
for interacting fields which does address the problems outlined above.  
The theory is similar in style to the conventional stochastic inflation 
program but differs in a fundamental way. In this theory we no longer 
identify the quantum fluctuations $\langle\psi^2\rangle$ directly
with the classical fluctuations that generate primordial density fluctuations.
The new role of the field $\psi$ is to
provide a noise source (via backreaction) in the quantum dynamics
of the local order parameter $\phi$. This is nothing but an application 
of the well known quantum Brownian motion paradigm 
of non-equilibrium statistical physics (see \cite{humat} 
and references therein). 
The field $\psi$ plays the role of an environment 
which couples to the system $\phi$ and indirectly generates fluctuations 
$\delta\phi$ in the system. This environmental noise will generate quantum
decoherence which is the process that leads to entropy generation and
the quantum-to-classical transition of the order parameter and its fluctuations.  
We then identify the resulting classical fluctuations of the
local order parameter $\delta\phi$, as those which lead to
density fluctuations, rather than the quantum fluctuations derived
directly from $\langle\psi^2\rangle$. Clearly in this approach the interaction 
between $\phi$ and $\psi$ is critical. 
As well as addressing the
quantum-to-classical transition problem, this theory leads to a dramatic 
easing of the fine tuning constraints, a problem that has plagued the 
conventional approach to inflaton dynamics.

In the classical limit of this theory, the dynamics of the local 
order parameter is described 
by \cite{matacz}
\begin{equation}
\ddot{\phi}+3H\dot{\phi}+V'(\phi)=\frac{H^2}{8\pi^3}V'''(\phi)F_c(t)
\end{equation}
where $F_c$ is a colored gaussian noise of unit amplitude with a 
correlation time of the order $H^{-1}$. 
The origin of the noise is the backreaction of
quantum fluctuations with wavelengths shorter than the
coarse-graining scale. 
The noise correlation function is ultraviolet {\em finite} and also 
turns out to be independent of any ultraviolet cutoff.
The noise is of a multiplicative nature
because its origin is the mode-mode coupling
induced by the self-interaction of the inflaton. For a free
field the stochastic term vanishes. This is because 
the environment $\psi$ and the system $\phi$ now decouple 
and the conventional situation is recovered. 
Significant simplification of (5) was obtained by
invoking the standard slow roll assumptions. This made it
possible to show that neglecting the potential
renormalization and non-local dissipation terms 
was a good first order approximation in the early slow roll phase.

The coarse-graining scale must be greater than the Hubble 
radius. This condition allows us to ignore the spatial gradient 
term in the system sector.
The theory is essentially independent of the coarse-graining scale
when this condition is met.
This robustness to the nature of the coarse-graining is an 
important virtue of the theory.
We also ignore information about
spatial correlations between the order parameters of different
regions. This allows a description based on a single degree of
freedom. 
Using the slow roll assumptions we can drop the inertial term in 
(5) and approximate the colored noise by a white noise. 
In this case we find that (5) becomes
\begin{equation}
\dot{\phi}+\frac{V'(\phi)}{3H}
=\frac{H^{1/2}}{\sqrt{864}\pi^3}V'''(\phi)F_w(t).
\end{equation}
$F_w$ is a white noise of unit amplitude that is interpreted in the 
Stratonovich sense (since it is an approximation to a colored noise).
We also interpret the noise in (4) the same way, though in this case 
one is also free to use the Ito interpretation. Equation (6) is the 
result we will use in this letter.

Equation (6) is valid only as long as the slow-roll approximation is valid. 
This approximation is valid when the slow-roll 
parameters $\epsilon$ and 
$|\eta|$ \cite{kinney}, which are defined by 
\begin{equation}
\epsilon(\phi)=\frac{m_{pl}^2}{16\pi}\left(\frac{V'(\phi)}{V(\phi)}\right)^2
\end{equation}
and
\begin{equation}
\eta(\phi)=\frac{m_{pl}^2}{8\pi}\left[\frac{V''(\phi)}{V(\phi)}-
\frac{1}{2}\left(\frac{V'(\phi)}{V(\phi)}\right)^2\right],
\end{equation}
are both less than 1.
During inflation we have $\epsilon(\phi)<1$, and the local 
order parameter rolls down the potential hill according to (2) from its 
initial value $\phi_0$. Inflation ends at a field value $\phi_e$ 
which is determined by
\begin{equation}
\epsilon(\phi_e)=\frac{m_{pl}^2}{16\pi}
\left(\frac{V'(\phi_e)}{V(\phi_e)}\right)^2=1.
\end{equation}
At this point slow-rolling ends and the reheating phase commences. 
Our discussion in this letter will be restricited to the potential  
\begin{equation}
V(\phi)=\lambda\phi^4/4
\end{equation}
for which equation (6) will be exactly solvable. 
For the potential (10) we find from 
(9) that $\phi_e=0.56m_{pl}$. This sets a lower bound on $\phi$.
The second slow roll condition is always satisfied up to this 
value.  

The number of e-folds of inflation which occur when the field evolves 
from $\phi$ to $\phi_e$ is \cite{inflation}
\begin{equation}
N(\phi)=\frac{8\pi}{m_{pl}^2}\int_{\phi_e}^{\phi}
\frac{V(\phi')}{V'(\phi')}d\phi'.
\end{equation}
Smoothness on scales comparable to the current observable universe requires
$N\ge 60$. This places a lower limit on the initial field value 
$\phi_0\ge \phi_{60}$, where $N(\phi_{60})=60$. For the potential (10) 
we find from (11) that
\begin{equation}
N(\phi)=\frac{\pi}{m_{pl}^2}(\phi^2-\phi_e^2).
\end{equation}
From this we can deduce $\phi_{60}=4.4m_{pl}$. 
In this letter, as is common in most inflation models, we 
assume that the observable universe leaves the horizon during inflation at
$60$ Hubble times before the end of inflation, i.e when the inflaton field 
has the value $\phi_{60}$.
The smallest scale for which density fluctuations can be probed $(\sim 1Mpc)$
will leave the horizon about $10$ Hubble times after the observable universe.
From (12) we find this corresponds to the field value $\phi_{50}=4.0m_{pl}$.
Significant is that the critical 
field values $\phi_e$ and $\phi_{60-50}$ are independent of $\lambda$.
The initial field $\phi_0$ cannot be arbitrarily large. It
must also satisfy $\phi_0\le \phi_{pl}$ where 
$\phi_{pl}$ is the field at the Planck boundary which satisfies
$V(\phi_{pl})= m_{pl}^4$. While $\phi_e$ and $\phi_{50-60}$ are independent of
$\lambda$, this is clearly not so for $\phi_{pl}$. 

In this letter we wish to calculate the statistical properties of 
observable density fluctuations predicted by (6) for the potential (10). 
Our aim is to demonstrate that this new approach to inflaton 
dynamics is consistent with the observed near gaussian and scale invariant 
density fluctuations.
It is well known from gauge-invariant analysis that the amplitude 
of a density fluctuation that crosses back inside the horizon
after inflation, can be deduced from the quantity \cite{inflation,fluct}
\begin{equation}
\frac{\delta \rho}{\rho}=\delta\phi\frac{H(\phi)}{\dot{\phi}},
\end{equation}
evaluated at the time the fluctuation scale of interest crossed 
outside the horizon during inflation. 
The power spectrum $\Delta(k)$ is related to the mean square density 
fluctuations $\delta\rho/\rho$ via  
\begin{equation}
\left(\frac{\delta \rho}{\rho}\right)^2=
\int^{\infty}_{-\infty}\Delta(k)d\ln k
\end{equation}
from which we obtain
\begin{equation}
\Delta(k)=\frac{d}{d\ln k}\left(\frac{\delta \rho}{\rho}\right)^2.
\end{equation}
The power spectrum is just the contribution, at a given time, 
to the mean square fluctuations generated in a Hubble time.
We are interested in the power spectrum over observable scales. 
Therefore we will evaluate the RHS of (15) at $\phi_{60}$
which is the classical field value at the time the scale of the 
observable universe leaves the horizon during inflation.
It is usual to assume that the density fluctuation power spectrum 
$\Delta (k)$ can, within the observable range of $k$, be written as 
\begin{equation}
\Delta(k)=A k^{n-1}
\end{equation}
where $A$ and $n$ are the amplitude and spectral index of the density
fluctuations. For $n=1$ we have a scale-invariant 
power spectrum of density fluctuations. 
From (16) we find that the spectral index can be determined 
from the power spectrum by
\begin{equation}
n=1+\frac{1}{\Delta(k)}\frac{d \Delta(k)}{d\ln k}.
\end{equation}

Clearly our first task is to calculate (13) for which we need to solve 
(6) in order to obtain $\delta\phi$.
As is commonly done in studies of the Starobinsky equation (4)
\cite{sifluct,yvm}, we 
include the effects of backreaction by simply assuming that 
$H$ is slowly varying as 
\begin{equation}
H^2(\phi)=\frac{8\pi}{3m_{pl}^2}V(\phi).
\end{equation}
This is possible since $H$ changes little over a Hubble time due to 
the slow rolling of the inflaton.
Upon substituting (10) and (18) into (6), and changing to the 
dimensionless variables $x=\phi/m_{pl}$ and $\tau= tm_{pl}$, we obtain
\begin{equation}
dx=-fxd\tau+gx^2F_w(\tau)d\tau
\end{equation}
where
\begin{equation}
f=\sqrt{\frac{\lambda}{6\pi}},\;\;\;g=\frac{1}{\sqrt{24}\pi^3}
\left(\frac{2\pi}{3}\right)^{1/4}\lambda^{5/4}.
\end{equation}
With the new variable 
$\chi=-1/x$ we find that (19) becomes
\begin{equation}
d\chi=f\chi+gF_w(\tau).
\end{equation}
This equation now describes an Ornstein-Uhlenbeck process for which 
the solution is 
\begin{equation}
\chi(\tau)=\chi_0e^{f\tau}+ge^{f\tau}\int^{\tau}_0e^{-fs}F_w(s)ds.
\end{equation}
Therefore our solution to (19) is 
\begin{equation}
x(\tau)=\frac{x_c(\tau)}{1-gx_0\int^{\tau}_0e^{-fs}F_w(s)ds},
\end{equation}
where the classical deterministic solution $x_c(\tau)$ is
\begin{equation}
x_c(\tau)=x_0e^{-f\tau}
\end{equation}
with $x_0$ as the initial field value. 
To obtain the fluctuations $\delta \phi$ we make a gaussian approximation 
\begin{equation}
x(\tau)\simeq x_c(\tau)\left(1+gx_0\int^{\tau}_0e^{-fs}F_w(s)ds\right)
\end{equation}
to (23) which will be justified later.
From this we obtain
\begin{eqnarray}
(\delta x)^2&=&\langle(x(\tau)-x_c(\tau))^2\rangle  \nonumber \\ 
&\simeq&\frac{\lambda^2}{24\pi^5}x_c^2(\tau)\bigl(x_0^2-x_c^2(\tau)\bigr).
\end{eqnarray}

We can compare this result directly to that predicted by the 
Starobinsky equation (4). Upon substituting (10) and (18) into (4), we find 
the exact solution to the Starobinsky equation is 
\begin{eqnarray}
x(\tau)&=&\frac{x_c(\tau)}{\sqrt{1-2hx_0^2\int^{\tau}_0e^{-2fs}F_w(s)ds}} 
\nonumber \\
&\simeq& \frac{x_c(\tau)}{1-hx_0^2\int^{\tau}_0e^{-2fs}F_w(s)ds},
\end{eqnarray}
where
\begin{equation}
h=\frac{1}{2\pi}\left(\frac{2\pi\lambda}{3}\right)^{3/4}.
\end{equation}
Making the same gaussian approximation as previously we find that 
\begin{equation}
(\delta x_s)^2\simeq\frac{\lambda}{12}
x_c^2(\tau)\bigl(x_0^4-x_c^4(\tau)\bigr)
\end{equation}
where we use the $s$ subscript to denote quantities evaluated from the 
solution to the Starobinsky equation. 

Using (26),(18) and (24) we find that (13) becomes
\begin{equation}
\left(\frac{\delta \rho}{\rho}\right)^2=\frac{1}{6\pi^3}\lambda^2 x_c^4(\tau)
\bigl(x_0^2-x_c^2(\tau)\bigr).
\end{equation}
To calculate the spectrum (15) we need to relate the classical field value 
at some time to the scale $k$ that is crossing the horizon at the same time.
We do this by first considering the number of e-folds $N(k)$
between the horizon crossing of a scale $k$ and the end of inflation. 
We are assuming that the scale of the observable universe left the 
horizon 60 Hubble times before the end of inflation. We can therefore write 
$N(k)$ as \cite{inflation} 
\begin{equation}
N(k)=60+\ln (k_*/k) 
\end{equation}
where $k_*$ is the scale of the current observable universe.
From (31) and (12) we find that 
\begin{equation}
x_c^2(\tau)=\frac{1}{\pi}\bigl[60+\ln (k_*/k)\bigr]+x_e^2.
\end{equation}
This determines the value of the field at the time a scale $k$ crosses 
the horizon. 
This is the result we will use to calculate the spectrum. 
Substituting (30) into (15) and using (32) we find
\begin{equation}
\Delta(k)=\frac{\lambda^2}{6\pi^4}x_c^2(\tau)\bigl(3x_c^2(\tau)-2x_0^2\bigr).
\end{equation}
The explicit $k$ dependence of this function is contained in $x_c(\tau)$
via equation (32). The spectrum has only a very weak logarithmic $k$ 
dependence. This will clearly give rise to a near scale-invariant spectrum. 
We can calculate the spectral index from (33) by using (17) and setting 
$x_c$ and $x_0$ to $x_{60}$. Observable 
constraints on the amplitude of density fluctuations require setting 
the LHS of (33) to $10^{-10}$. Setting $x_c$ and $x_0$ to $x_{60}$ allows 
us to calculate a value for $\lambda$. The results are 
\begin{equation}
\lambda=1.2\times 10^{-5},\;\;\;n=0.93.
\end{equation}
We can perform an exactly analogous calculation using the fluctuations 
(29) derived from the exact solution of the Starobinsky equation. We find 
the familiar result
\begin{equation}
\lambda_s=6.6\times 10^{-15},\;\;\;n_s=0.92.
\end{equation}
We see that equation (6) leads to a dramatic easing of the fine-tuning 
required, but still gives a near scale-invariant spectrum of density
fluctuations. Similar easings of the fine-tuning constraints have been 
reported in \cite{morwas,CH1,calgon,matacz}.

Also of great interest is a measure of how much the probability distribution 
of fluctuations deviates from gaussian. 
When the magnitude of the stochastic 
term in the denominator of (23) is small ($\ll 1$), it becomes possible 
to make the gaussian approximation (25) to the exact solution (23).  
We are therefore led to define
\begin{equation}
d=\frac{gx_0}{\sqrt{2f}}=\frac{x_0\lambda}{\sqrt{24\pi^5}}
\end{equation}
as a simple measure of the deviation from gaussian. (The denominator 
$\sqrt{2f}$ of (36) is the typical fluctuation size of the 
stochastic term in (23) in the stationary limit.) 
Using $x_0=4.4$, and the result (34) for 
$\lambda$, we find $d=6\times 10^{-7}$.
We can compare (36) to that derived from the 
exact solution (27) of the Starobinsky equation.
In this case we find 
\begin{equation}
d_s=\frac{hx_0^2}{2f^{1/2}}=\left(\frac{\lambda_s}{12}\right)^{1/2}x_0^2.
\end{equation}
Using the result (35) for $\lambda_s$, and $x_0=4.4$,
we find $d_s=5\times 10^{-7}$. We therefore see that over scales of the 
observable universe both solutions predict nearly identical and negligable 
deviations from gaussian. 

By assuming $x_0=x_{60}$ we are simply trying to exclude the effects of 
fluctuations on scales larger than our present horizon. In actual fact 
inflation would most likely have been in progress for a long time before the 
observable universe left the Hubble radius during inflation.  
In this case we can consider $x_0$ to be anywhere in the range 
$4.4\le x_0\le x_{pl}$. For the usual self-coupling constant in (35) 
we find that the Planck boundary is $x_{pl}\simeq 5000$. Because 
the deviations from gaussian in (37) are proportional to 
$x_0^2$ there is a possibility of significant deviations from 
gaussian fluctuations on super-horizon scales.
For the new coupling constant in (34) we have $x_{pl}\simeq 24$. 
In this case the deviations from gaussian in (36) depend only linearly on 
$x_0$. This combined with the much smaller Planck boundary means that 
the solution (23) predicts gaussian fluctuations to a high accuracy over 
both super and sub-horizon scales. We have ignored the effects 
of boundary conditions that should be imposed at $\phi_e$ and $\phi_{pl}$. 
These effects will be small and and are unlikely to change these
conclusions.

Yi, Vishniac and Minehige \cite{yvm} have analysed 
in detail solutions of the Starobinsky equation (in both Ito and 
Statonovich interpretations) for the model discussed here. They discussed 
deviations from gaussian in terms of a simple measure for skewness derived 
from the probability distribution of the exact solution (27). We have 
also applied this measure to the exact solution (23) and found it to give
results consistent with the simpler measure discussed above. 

In this letter we showed that the new theory of inflaton 
dynamics developed in \cite{matacz} can address the fine-tuning problem
yet still predict density fluctuations in general agreement with observations.
This was demonstrated for the quartic potential but is likely to generalize 
to a wide range of potentials.
These results were based on the classical limit of the 
theory. The other great advantage of this theory is that it leads naturally to 
a description of the inflaton as a quantum open system. This allows the
quantum-to-classical transition to occur as a non-equilibrium quantum 
statistical process (decoherence), rather than being simply postulated as 
in the conventional approach. Work on this issue is currently in progress.

\acknowledgments
I would like to thank the Australia Research Council for their
generous support of this research
through an Australian Postdoctoral Research Fellowship and an ARC
small grant.


\begin{references}
\bibitem{inflation} T. Padmanabhan, {\it Structure Formation in the
Early Universe}, Cambridge University Press (1993);
P.J.E. Peebles, {\it Principles of Physical Cosmology},
Princeton University Press, N.J. (1993);
A.R. Liddle and D.H. Lyth, Phys. Rep. {\bf 231}, 1 (1993);
E.W. Kolb and M.S. Turner, {\it The Early Universe}, Addison Wesley (1990).

\bibitem{vfl}
A. Vilenkin and L. Ford, Phys. Rev. {\bf D26}, 1231 (1982);
A. Linde, Phys. Lett. {\bf 116B}, 335 (1982).

\bibitem {fluct}
A. Guth and S. Y. Pi, Phys. Rev. Lett. {\bf 49}, 1110 (1982);
A. A. Starobinsky, Phys. Lett. {\bf 117B}, 175 (1982);
S. W. Hawking, Phys. Lett. {\bf 115B}, 295 (1982);
J. M. Bardeen, P. J. Steinhardt and M. S. Turner, Phys. Rev.
{\bf D28}, 629 (1983);
R. Brandenberger, R. Kahn and W. Press, Phys. Rev. {\bf D 28}, 1809
(1983);
V. Mukhanov, H. Feldman and R. Brandenberger, Phys. Rep. {\bf 215},
203 (1992);
R. Brandenberger, Nucl. Phys. B. {\bf 245}, 328 (1984);
A. H. Guth and S. Y. Pi, Phys. Rev. {\bf D32}, 1899 (1985).


\bibitem {staro}
A. A. Starobinsky, in {\it Field Theory, Quantum Gravity and
Strings}, ed. H. J. de Vega and N. Sanchez (Springer, Berlin 1986).
                                                                     
\bibitem{stoinf}
S. J. Rey, Nucl. Phys. {\bf B284}, 706 (1987);
J. M. Bardeen and G. J. Bublik, Class. Quan. Grav. {\bf 4}, 473
(1987); M. Morikawa, Phys. Rev. {\bf D42}, 1027
(1990); H. E. Kandrup, Phys. Rev. {\bf 39}, 2245 (1989);
A. Hosoya, M. Morikawa and K. Nakayama,  Int. J. Mod. Phys. A
{\bf 4}, 2613 (1989);
Y. Nambu, Phys. Lett. B {\bf 276}, 11 (1992); M. Mijic,
Phys. Rev. {\bf D49}, 6434 (1994); 
F.R. Graziani, Phys. Rev. {\bf D38}, 1122 (1988);{\bf D38}, 1131
(1988);{\bf D38}, 1802 (1988); {\bf D39}, 3630
(1989); O.E. Buryak, Phys. Rev. D {\bf 53}, 1763 (1996).

\bibitem{sifluct}
I. Yi and E. T. Vishniac, Phys. Rev. {\bf D43}, 5295 (1993);
Phys. Rev. {\bf D45}, 3441 (1992);
D.S. Salopek and J.R. Bond, Phys. Rev. {\bf D43}, 1005 (1991);
S. Mollerach, S. Matarrese, A. Ortolan, and
F. Lucchin, Phys. Rev. {\bf D44}, 1670 (1991);
G.V. Chibisov and Yu. V. Shtanov, Int. J. Mod. Phys. {\bf A13}, 2625
(1990);
D.S. Salopek, J.R. Bond and J.M. Bardeen, Phys. Rev. {\bf D40}, 1753
(1989); S. Matarrese, A. Ortolan and F. Lucchin, Phys. Rev.
{\bf D40}, 290 (1989); A. Ortolan, F. Lucchin and S. Matarrese,
Phys. Rev. {\bf D38}, 462 (1988). 

\bibitem{yvm}
I. Yi, E.T. Vishniac and S. Mineshige, Phys. Rev. {\bf D43}, 362
(1991).



\bibitem{vlss}
A. Linde, D. Linde and A. Mezhlumian, Phys. Rev. D {\bf 49}, 1783
(1994); Phys. Lett. {\bf B307}, 25 (1993); Phys. Lett. {\bf B345},
203 (1995);
Y. Nambu and M. Sasaki, Phys. Lett. B {\bf 219}, 240 (1989);
Y. Nambu, Prog. Theor. Phys {\bf 81}, 1037 (1989).


\bibitem{cgea}
B. L. Hu and Y. Zhang, ``Coarse-Graining, Scaling, and
Inflation'' Univ. Maryland Preprint 90-186 (1990); B. L. Hu,
in {\it Relativity and Gravitation: Classical and Quantum}
Proceedings of SILARG VII, Cocoyoc, Mexico, Dec. 1990.
eds. J. C. D' Olivo {\it et al} (World Scientific, Singapore, 1991).

\bibitem{HuBelgium}
B. L. Hu, J. P. Paz and Y. Zhang, in {\it The Origin of Structure in
the Universe}, ed. by E. Gunzig and
P. Nardone (Kluwer, Dordrecht 1993);
B. L. Hu, in {\it Quantum Physics and the Universe}, ed. Namiki,
K. Maeda, et al  (Pergamon Press, Tokyo, 1993).
Vistas in Astronomy {\bf 37}, 391 (1993); 


\bibitem{sim}
F. Lombardo and F.D. Mazzitelli, Phys. Rev. D {\bf 53},
2001 (1996).

\bibitem {mm}
M. Morikawa, Prog. Theor. Phys {\bf 93}, 685 (1995).
                                       
\bibitem{morwas}
M.Morikawa, in {\it Quantum Physics and the Universe}, ed. Namiki,
K. Maeda, et al  (Pergamon Press, Tokyo, 1993).
Vistas in Astronomy, {\bf 37}, 87 (1993).

\bibitem{CH1}
E. Calzetta and B.L. Hu, Phys. Rev. D {\bf 52}, 6770 (1995).

\bibitem{calgon}
E. Calzetta and S. Gonorazky, `Primordial Fluctuations and Nonlinear
Couplings', gr-qc/9608057 (1996).

\bibitem{matacz}
A. Matacz, `A New Theory of Stochastic Inflation', gr-qc/9604022 (1996).

\bibitem{humat}
B.L. Hu and A. Matacz, Phys. Rev. D {\bf 49}, 6612 (1994).

\bibitem{kinney}
W. H. Kinney and K. T. Mahanthappa, Phys. Rev. D {\bf 53}, 5455 (1996).






\end{references}
\end{document}